\documentclass[twocolumn,noshowpacs,amsmath,amssymb]{revtex4-2}
\usepackage{graphicx}
\usepackage{amsfonts}
\usepackage{dcolumn}
\usepackage{bm}
\usepackage{color}
\usepackage{hyperref}
\begin{document}
\title{Large deviation and anomalous fluctuations scaling in degree assortativity on configuration networks}

\author{Hanshuang Chen$^{1}$}\email{chenhshf@ahu.edu.cn}

\author{Feng Huang$^{2}$}

\author{Chuansheng Shen$^{3}$}\email{csshen@mail.ustc.edu.cn}

\author{Guofeng Li$^{1}$}

\author{Haifeng Zhang$^{4}$}

\affiliation{$^{1}$School of Physics and Optoelectronics Engineering, Anhui
University, Hefei 230601, China \\$^{2}$School of Mathematics and
Physics, Anhui Jianzhu University, Hefei 230601, China
\\ $^{3}$School of Mathematics and
Physics, Anqing Normal University, Anqing 246133, China  \\
$^{4}$School of Mathematical Science, Anhui University, Hefei
230601, China}


\begin{abstract}
By constructing a multicanonical Monte Carlo simulation, we obtain
the full probability distribution $\rho_N(r)$ of the degree
assortativity coefficient $r$ on configuration networks of size $N$
by using the multiple histogram reweighting method. We suggest that
$\rho_N(r)$ obeys a large deviation principle, $\rho_N \left( r-
r_N^* \right) \asymp {e^{ - {N^\xi }I\left( {r- r_N^* } \right)}}$,
where the rate function $I$ is convex and possesses its unique
minimum at $r=r_N^*$, and $\xi$ is an exponent that scales
$\rho_N$'s with $N$. We show that $\xi=1$ for Poisson random graphs,
and $\xi\geq1$ for scale-free networks in which $\xi$ is a
decreasing function of the degree distribution exponent $\gamma$.
Our results reveal that the fluctuations of $r$ exhibits an
anomalous scaling with $N$ in highly heterogeneous networks.
\end{abstract}
 \maketitle

\section{Introduction}
Over the past two decades, we have witnessed the success of complex
networks in describing the pattern discovered ubiquitously in real
world \cite{Newman_NetworkBook}, such as community structure and
scale-free structure, and modelling many dynamical processes in
nature \cite{RMP08001275}, such as synchronization
\cite{PRP08000093,PRP2016}, epidemic spreading
\cite{RevModPhys.87.925}, opinion formation \cite{RMP09000591}, etc
\cite{PR.687.1}. \textcolor{blue}{In particular}, how to characterize the structural
features of complex networks is essential not only for uncovering
the organizational principles of real systems, but also for
understanding and controlling the dynamical processes on them
\cite{PRP06000175,PRP2016(2),RevModPhys.88.035006}.

An important feature in complex networks is so-called degree
assortativity, which quantifies the tendency of nodes to be
connected to other nodes of similar degree. A networks is called
assortative if nodes with high degree preferably connect to other
nodes with high degree, and dissortative if nodes with high degree
are linked to nodes with low degree. Technical and biological
networks have been found to be dissortatively mixed, while social
networks show assortative correlations
\cite{SIR03000167,PhysRevLett.89.208701,PhysRevE.67.026126}. It was
shown that, on the one hand, degree correlations are key to many
structural properties of networks, such as percolation
\cite{PhysRevLett.89.208701,PRE.70.06610}, mean distance
\cite{PRE.70.06610}, and robustness
\cite{PhysRevE.67.026126,PhysRevE.67.015101}. On the other hand,
degree correlations affect the properties of dynamical processes
taking place on networks, such as epidemic spreading
\cite{PhysRevLett.89.108701,PhysRevE.66.047104,PhysRevLett.90.028701},
stability against stimuli and perturbation
\cite{Sinha2005,PhysRevE.75.046113}, and synchronization of
oscillators \cite{PhysicaD224.123.2006}.

In his seminal papers
\cite{PhysRevLett.89.208701,PhysRevE.67.026126}, Newman introduced
the assortativity coefficient $r$ to measure the degree correlation,
which is defined as
\begin{eqnarray}
r = \frac{{{M^{ - 1}}\sum\nolimits_i {{j_i}{k_i} - {{\left[ {{M^{ -
1}}\sum\nolimits_i {\frac{1}{2}\left( {{j_i} + {k_i}} \right)} }
\right]}^2}} }}{{{M^{ - 1}}\sum\nolimits_i {\frac{1}{2}\left( {j_i^2
+ k_i^2} \right)}  - {{\left[ {{M^{ - 1}}\sum\nolimits_i
{\frac{1}{2}\left( {{j_i} + {k_i}} \right)} }
\right]}^2}}},\label{eq1}
\end{eqnarray}
where $M$ is the number of edges, and $j_i$, $k_i$ are the degrees
of the nodes at the ends of the $i$th edge, with $i = 1,\cdots,M$.
The assortativity coefficient $r$ is actually the Pearson's
correlation coefficient between the degrees of neighboring nodes,
which is supposed to have natural bounds $r \in [-1, 1]$. A network
is assortative when $r > 0$ and disassortative when $r < 0$.

Most of previous works on this subject were performed on scale-free networks
with power-law degree distributions $P(k)\sim k^{-\gamma}$
\cite{Maslov_PhysicaA_2004,PhysRevE.68.026112,PhysRevLett.104.108702,PhysRevE.81.046103,PhysRevE.81.031135,Chaos27.033113.2017,PhysRevE.82.037102,PhysRevE.87.022801}.
It has been shown that, on the one hand, for degree distribution
exponent $2<\gamma<4$ the assortativity coefficient $r$ is usually
negative in finite-size networks. On the other hand, $r$ always
decreases in magnitude as network size increases, and $r$ equals to
zero in the infinite networks. Maslov \emph{et al.}
\cite{Maslov_PhysicaA_2004} have shown by using computer simulations
that the degree dissortativity results from the restriction of at
most one edge between any pair of nodes. Furthermore, Park and
Newman \cite{PhysRevE.68.026112} verified this result in theory.
They proposed a grand canonical ensemble of graphs such that
analytical calculation of degree correlations becomes feasible.
Johnson \emph{et al.} \cite{PhysRevLett.104.108702} proposed an
alternative explanation for the phenomenon by information entropy,
and they showed that the Shannon entropy is maximized at some
negative value of assortativity coefficient $r$ for highly
heterogeneous scale-free networks. Menche \emph{et al.}
\cite{PhysRevE.81.046103} analyzed the maximally disassortative
scale-free networks and found that the lower bound of $r$ approaches
to zero as network size increases in a power-law way. Dorogovtsev
\emph{et al.} \cite{PhysRevE.81.031135} also found the results in a
specific class of recursive trees with power-law degree
distribution. Yang \emph{et al.} \cite{Chaos27.033113.2017} derived
analytically the lower bound of assortativity coefficient in
scale-free networks. Similar phenomenon was also discussed in some
related works \cite{PhysRevE.82.037102,PhysRevE.87.022801}, although
the authors therein argued the availability of the Pearson's
coefficient for measuring degree correlations in large-size
heavy-tailed networks, and alternatively they proposed other
measurements such as Kendall-Gibbons' $\tau_b$
\cite{PhysRevE.82.037102} and Spearman's $\rho$
\cite{PhysRevE.87.022801}.

Previous works mainly focused on either the typical behavior of $r$,
such as how the expected value of $r$ changes with network size and
degree heterogeneity
\cite{Maslov_PhysicaA_2004,PhysRevE.68.026112,PhysRevLett.104.108702,PhysRevE.81.031135},
or how to obtain a class of specific networks with some atypical
value of $r$
\cite{PRE.70.06610,PhysRevE.81.046103,Chaos27.033113.2017}. For an
ensemble of random networks with a given degree sequence (i.e.
configuration model), it is known that the assortativity coefficient
$r$ varies from one network realization to another. An interesting
question arises: what is the probability of generating a
configuration network whose assortativity coefficient $r$ falls in
an interval $[r, r+\mathrm{d}r)$? The question is equivalent to
finding the probability distribution function $\rho_N$ of $r$ with
network size $N$. For the purpose, we shall employ a
statistical-mechanics inspired Monte Carlo (MC) method, multiple
histogram reweighting (MHR)
\cite{PhysRevLett.61.2635,PhysRevLett.63.1658.2}, to fully sample
$\rho_N$ over a wide range of $r$. The method is \textcolor{blue}{computationally} 
efficient and enable us to cover rare-event tails with very low
probabilities of $r$. Recently, the MHR method was applied to
investigate the large deviation properties of the largest connected
\cite{Hartmann_EPJB_2011} or biconnected component
\cite{Hartmann_EPJB_2019}, the diameters \cite{PhysRevE.97.032128}
for random graphs, and resilience of transportation networks
\cite{Hartmann_EPJB_2014} as well as power grids
\cite{Hartmann_NJP_2015}. Related algorithms
\cite{Saito_Iba_Kitajima_2014}, for example, Wang-Landau algorithm
\cite{PhysRevLett.86.2050}, has been used to efficiently sample
large spectral gap \cite{Saito_Iba_2011} and prescribed motif
densities in networks \cite{PhysRevLett.115.188701}, and rare
trajectories in chaotic systems
\cite{Altmann_EPJB_2017,Altmann_Chaos_2019}.

To \textcolor{blue}{that} end, we first build a canonical ensemble MC sampling by a
random edge-swapping scheme \cite{Maslov_Sneppen_2002} and then
collect a series of histograms of $r$ at different inverse
temperatures. Finally, $\rho_N(r)$ is obtained by using the MHR
method.  By implementing the method on the configuration models with
Poisson degree distributions and power-law degree distributions, we
find that for all the cases under consideration $\rho_N(r)$ is
unimodal and its width becomes narrower as $N$ increases. The
expected value of $r$ is negative and decays in magnitude as $N$
increases in a power-law way, as reported in previous literatures.
The variance $\sigma_r^2$ of $r$ decreases in power-law form,
$\sigma_r^2 \propto 1/N^\xi$, with the increase of $N$ as well. For
homogeneous networks such as Poisson random graphs, $\xi=1$ such
that the fluctuation in $r$ is standard. Strikingly, for highly
heterogeneous networks such as scale-free networks with $\gamma<3$,
we have $\xi>1$ and thus the fluctuation scaling of $r$ with $N$ is
\textcolor{blue}{anomalous}. Moreover, we suggest that $\rho_N(r)$ obeys a large
deviation principle \cite{Touchette_PhysRep_2009}, $\rho_N \left( r-
r_N^* \right) \asymp {e^{ - {N^\xi }I\left( {r- r_N^* } \right)}}$,
where ${I\left( {r } \right)}$ is so-called large deviation rate function which plays a role of microcanonical entropy of the network configuration model \cite{Piraveenan_2009,PhysRevE.80.045102,PhysRevE.82.011116,PhysRevE.89.062807}.
$r_N^*$ is the most probable value of $r$, and $\xi$ is just
mentioned that is the exponent scaling the $\sigma_r^2$'s with $N$.

\section{Multi-Canonical ensemble Monte Carlo sampling}
The configuration model is an ensemble of random graphs with a given
degree sequence $\{ k_1,\cdots,k_N \}$, where $k_i$ is the
degree of node $i$ and $N$ is the number of nodes. The model was
formulated by Bollob\'as \cite{Bollobas1980}, inspired by
Ref.\cite{Bender1978}. It was popularized by Newman, Strogatz, and
Watts \cite{PhysRevE.64.026118}, who realized that it is a useful
and simple model for real-world networks. The configurations
networks are generated as follows. Firstly, each node $i$ is
assigned a given number of half-edges equal to its observed degree
$k_i$, with $\sum\nolimits_{i = 1}^N {k_i}$ assumed to be even. Each
half-edge is then connected to a randomly chosen other half-edge to
form an edge in the graph. Finally, all the self-loops and all the
parallel edges between two different nodes are removed by an
algorithm to reshuffle edges that ensures the degree distribution
unchanged. \textcolor{blue}{It was pointed out that the algorithm produces a bias in resulting network configurations \cite{PhysRevE.85.026101}. Such a bias can be eliminated by a refusal algorithm \cite{Britton2006}, but the latter is more computationally time-consuming. However, it does not produce any effect in our model whether algorithm is applied. This is because that the first generated network is only used as the starting point of Monte Carlo sampling introduced below. In the long time, the results do not sensitive to the initial configuration.}

\begin{figure}
	\centerline{\includegraphics*[width=1.0\columnwidth]{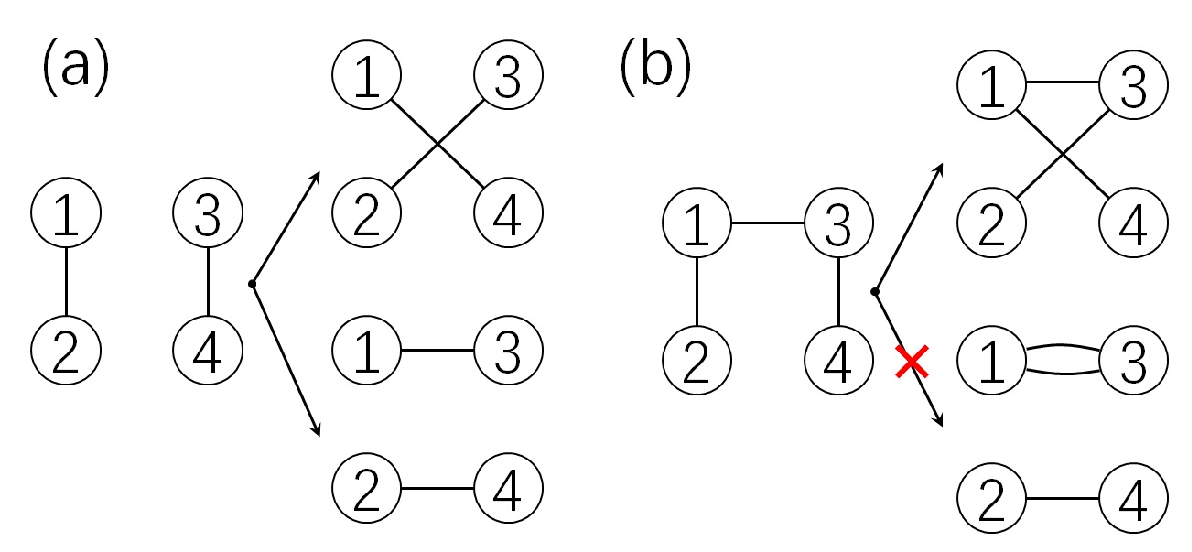}}
	\caption{(color online). \textcolor{blue}{Examples of edge swaps in two simple networks of four nodes with two edges (a) or three edges (b). For (a), there are two candidate configurations to move via edge swaps. While for (b), there remains only one move configuration, since the other is forbidden due to the presence of parallel edges.}    \label{fignet}}
\end{figure}

We consider a Markov Chain Monte Carlo  (MCMC) algorithm in which we weight each network configuration ${\bf A}$ with a Boltzmann weight $r({\bf A})$, where \textcolor{blue}{${\bf A}$ is the adjacency matrix of the underlying network whose entries are defined as $A_{ij}=A_{ji}=1$ if nodes $i$ and $j$ are connected and $A_{ij}=A_{ji}=0$ otherwise}, and $r({\bf A})$ is the assortativity coefficient of the network ${\bf A}$. To perform the MCMC, we consider the elementary edge swap moves that preserve the degree distribution of the network.
We consider four different nodes, $i,j,k,\ell$, 
and one of the three invertible moves ${\bf A}\to F{\bf A}$ in which the following edge swaps are performed \cite{coolen2009constrained,Coolen2017}
\begin{subequations}
\begin{align} 	
F_{ijk\ell;[1]}: \quad (i,j) &(k,\ell)\leftrightarrow(j,k)(i,\ell), \\
F_{ijk\ell;[2]}: \quad (i,j)&(k,\ell)\leftrightarrow (j,\ell)(i,k), \\
F_{ijk\ell;[3]}: \quad (i,k)&(j,\ell)\leftrightarrow (j,k)(i,\ell).
\end{align}
\end{subequations}
In order to perform any of these three moves the initial two links between the four nodes must be present in the network while the final two links must be absent or vice versa as multiple edges between two
different vertices are forbidden. 
Since not all moves are accepted by the algorithm the MCMC algorithm  should take into account the fact that some network configurations might allow more moves than others. \textcolor{blue}{In Fig. \ref{fignet}(a), we show a simple graph of four nodes with two edges. There exist two possible configurations to move by edge swaps. However, if an additional edge is introduced between nodes 1 and 3 (see Fig. \ref{fignet}(b)), one of the resulting move configurations is forbidden since the parallel edges are present.}

We indicate with $|\Phi_{\bf A}|$ the number of edge swaps allowed if starting from adjacency matrix ${\bf A}$.
Each single {\em allowable} edge-swap move ${\bf A}\to F{\bf A}$ is accepted by the Metropolis probability which ensure unbiased sampling of the network configurations
\begin{eqnarray}
P_{acc}=\min \left\{ {1,e^{  { - \beta \Delta r} }\frac{|\Phi_{\bf A}|}{|\Phi_{F{\bf A}}|}}
\right\},\label{eq2}
\end{eqnarray}
where $\beta$ is the inverse temperature \textcolor{blue}{played a role of a conjugated field acting on the assortativity coefficient $r$. Generally, for a larger $\beta$, the sampled networks prefers to smaller values of $r$, and thus $\beta$ can be used to adjust the bias on sampling assortativity coefficient.} $\Delta r$ is the
change in the assortativity coefficient $r$ due to the edge-swapping
trial. 
Therefore at each step  the algorithm \textcolor{blue}{selects} a value  of $\alpha\in\{1,2,3\}$ with uniform probability and draws four nodes $i<j<k<\ell$ until the move $F_{ijk\ell;[\alpha]}$ is allowed.
It then accepts the allowed move with probability $P_{acc}$.

We note that $|\Phi_{\bf A}|$ admits the following expression
\begin{eqnarray}
|\Phi_{\bf A}|&=&\frac{1}{4}\left(\sum_{i=1}^Nk_i\right)^2+\frac{1}{4}\sum_{i=1}^Nk_i-\frac{1}{2}\sum_{i=1}^N k_i^2\nonumber \\
&&-\frac{1}{2}\sum_{i,j}A_{ij}k_ik_j+\frac{1}{2}\mbox{Tr} ({\bf A}^3)+\frac{1}{4}\mbox{Tr}({\bf A}^4).
\end{eqnarray}
This expression can be used to calculate $|\Phi_{\bf A}|$ at the beginning of the MCMC algorithm.
In order to calculate how $|\Phi_{\bf A}|$ changes at each step of the MonteCarlo step it is more convenient to consider the expression 
\begin{eqnarray}
|\Phi_{\bf A}|&=& \sum_{i<j<k<\ell}\left[A_{ij}A_{k\ell}(1-A_{jk})(1-A_{i\ell})\right.\nonumber \\
&&+A_{ij}A_{k\ell}(1-A_{ik})(1-A_{j\ell})\nonumber \\
&&+A_{ik}A_{j\ell}(1-A_{jk})(1-A_{i\ell})\nonumber \\
&&+(1-A_{ij})(1-A_{k\ell})A_{jk}A_{i\ell}\nonumber \\
&&+(1-A_{ij})(1-A_{k\ell})A_{ik}A_{j\ell}\nonumber \\
&&\left.+(1-A_{ik})(1-A_{j\ell})A_{jk}A_{i\ell}\right].\label{Aexp}
\end{eqnarray}
Indeed using this expression one can just write
\begin{eqnarray}
|\Phi_{F{\bf A}}|=|\Phi_{\bf A}|+\Delta|\Phi|,
\end{eqnarray}
where $\Delta |\Phi|$ can be calculated by considering only the terms that change in Eq. (\ref{Aexp})

In fact, the term ${|\Phi_{\bf A}|}/{|\Phi_{F{\bf A}}|}$ in Eq.(\ref{eq2}) is very close to one since ${|\Phi_{\bf A}|}$ is the order of square of the number of edges, $M^2$, and thus the deviation of ${|\Phi_{\bf A}|}/{|\Phi_{F{\bf A}}|}$ from one is the order of $1/M^2$. Therefore, dropping such a term is expected to generate not much effect to the results, but it is bound to improve computing efficiency significantly. We have tested several networks and found that the results are consistent whether the term exists or not.  

Similar procedure was also used to study the relation between degree correlations and other topological features
such as clustering coefficient \cite{Ramos_Anteneodo_2013} and
percolation property \cite{PhysRevE.86.066103}. For a given inverse
temperature $\beta_i$, the probability density $p_i(r)$ of
generating a network with the assortativity coefficient $r$ follows
the Boltzmann distribution
\cite{PhysRevE.70.066117,Garlaschelli_NJP2011,Garlaschelli_NatRevPhys2019},
\begin{eqnarray}
{p_i}\left( r \right) = \rho_N \left( r \right)\frac{{{e^{ - {\beta
_i}r}}}}{{{Z_i}}},\label{eq3}
\end{eqnarray}
where $\rho_N( r )$ is probability density function of $r$ we want
to obtain, and ${Z_i} = \int {\rho_N \left( r \right){e^{ - {\beta
_i}r}}\mathrm{d}r} $ is the partition function (normalized factor)
at the inverse temperature $\beta_i$. In practice, $p_i$ can be
obtained by performing MC simulations at $\beta_i$. To \textcolor{blue}{that} end, we
build a histogram $N_i(r)$ of the number of times out of $n_i$ that
an interval $\left[r, r+\mathrm{d}r\right)$ is observed, and thus we
have
\begin{eqnarray}
{p_i}\left( r \right)\mathrm{d}r =
\frac{{{N_i}(r)}}{{{n_i}}}.\label{eq4}
\end{eqnarray}
\textcolor{blue}{In simulations, we have performed $N \times 10^5$ (with $N$ being the size of the underlying network) trials for edge swaps and the last $5N \times 10^4$ trials are used to count bins of histogram of $r$.} Using Eq. (\ref{eq4}), Eq. (\ref{eq3}) can be rewritten as
\begin{eqnarray}
\rho_N \left( r \right)\mathrm{d}r =
\frac{{{N_i}(r){Z_i}}}{{{n_i}{e^{ - {\beta _i}r}}}}.\label{eq5}
\end{eqnarray}
The MHR method takes advantage of collecting a series of histograms
at nearby temperature overlap. We perform a series of $R$ MC
simulations in the canonical ensemble corresponding to $R$ different
inverse temperature $\beta_i$ with $i=1,\cdots,R$, where $\beta_i$
is chosen uniformly from the interval $\left[\beta_{min},
\beta_{max}\right]$. The improved estimate for $\rho_N(r)$ is given
by \cite{Newman_MCBook}
\begin{eqnarray}
\rho_N \left( r \right)\mathrm{d}r = \frac{{\sum\nolimits_{i = 1}^R
{{N_i}\left( r \right)} }}{{\sum\nolimits_{j = 1}^R {{n_j}Z_j^{ -
1}{e^{ - {\beta _j}r}}} }},\label{eq6}
\end{eqnarray}
where the partition function $Z_j$ can be found self-consistently by
iterating the following equations,
\begin{eqnarray}\label{eq7}
{Z_k} = \int_r {\rho_N \left( r \right){e^{ - {\beta
_k}r}}\mathrm{d}r} = \int_r {\frac{{\sum\nolimits_{i = 1}^R
{{N_i}\left( r \right)} }}{{\sum\nolimits_{j = 1}^R {{n_j}Z_j^{ -
1}{e^{\left( {{\beta _k} - {\beta _j}} \right)r}}} }}\mathrm{d}r}. \nonumber \\
\end{eqnarray}
\textcolor{blue}{During the iterations for Eq.(\ref{eq7}), we have used a rescaling of $Z$-values (divided all by the smallest) after each step to avoid an overall growth.}  

Once the $\rho_N(r)$ is obtained, we can compute the $n$th moment of
the assortativity coefficient $r$,
\begin{eqnarray}
\left\langle {{r^n}} \right\rangle  = \frac{{\int {{r^n}\rho \left(
r \right)\mathrm{d}r} }}{{\int {\rho \left( r \right)\mathrm{d}r} }}.\label{eq8}
\end{eqnarray}
In particular, $\left\langle r \right\rangle$ is the expected value
of $r$, and $\sigma _r^2 = \left\langle {{r^2}} \right\rangle  -
{\left\langle r \right\rangle ^2}$ is the variance of $r$.

\section{Poisson random graphs}

\begin{figure}
\centerline{\includegraphics*[width=0.8\columnwidth]{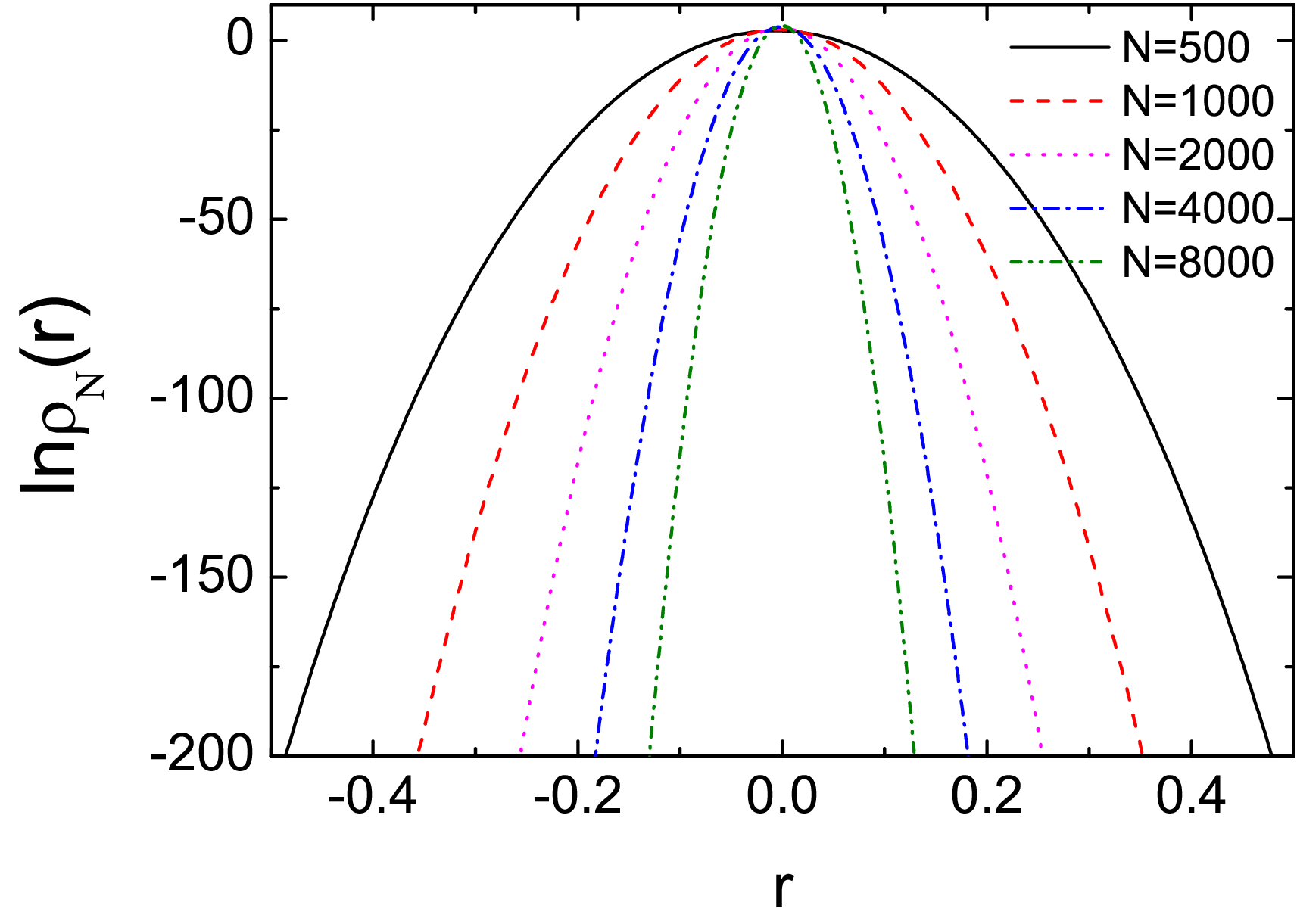}}
\caption{(color online). Logarithm of the probability distribution
$\rho_N(r)$ of the assortativity coefficient $r$ on Poisson random
graphs for different network size $N$ and a fixed average degree
$\left\langle k \right\rangle=6$. \label{fig1}}
\end{figure}

\begin{figure}
\centerline{\includegraphics*[width=1.0\columnwidth]{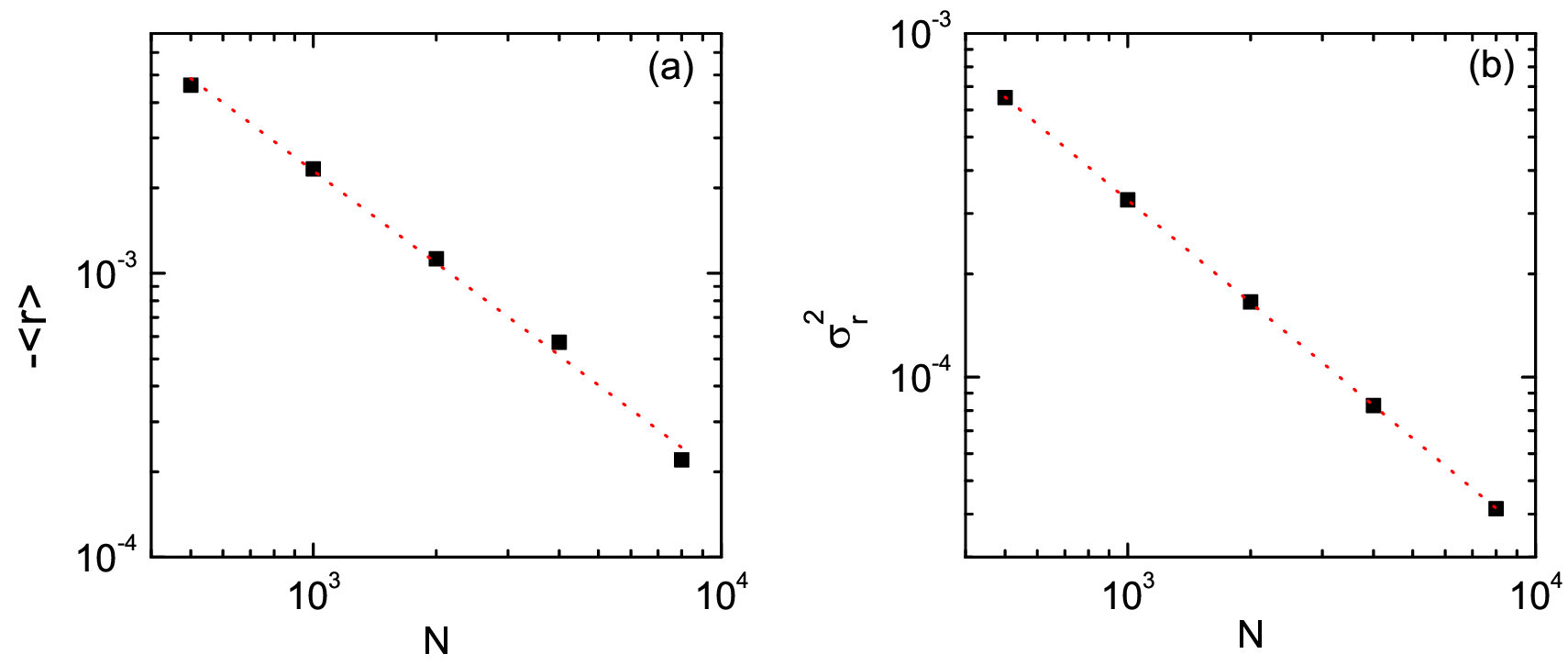}}
\caption{(color online). Log-log plot of the minus expected value
$-\left\langle r \right\rangle $ (a) and the variance $\sigma _r^2$
(b) of $r$ as a function of $N$ on Poisson random graphs with the
average degree $\left\langle k \right\rangle=6$. The dotted lines
show the linear fittings. \label{fig2}}
\end{figure}

\begin{figure}
\centerline{\includegraphics*[width=0.8\columnwidth]{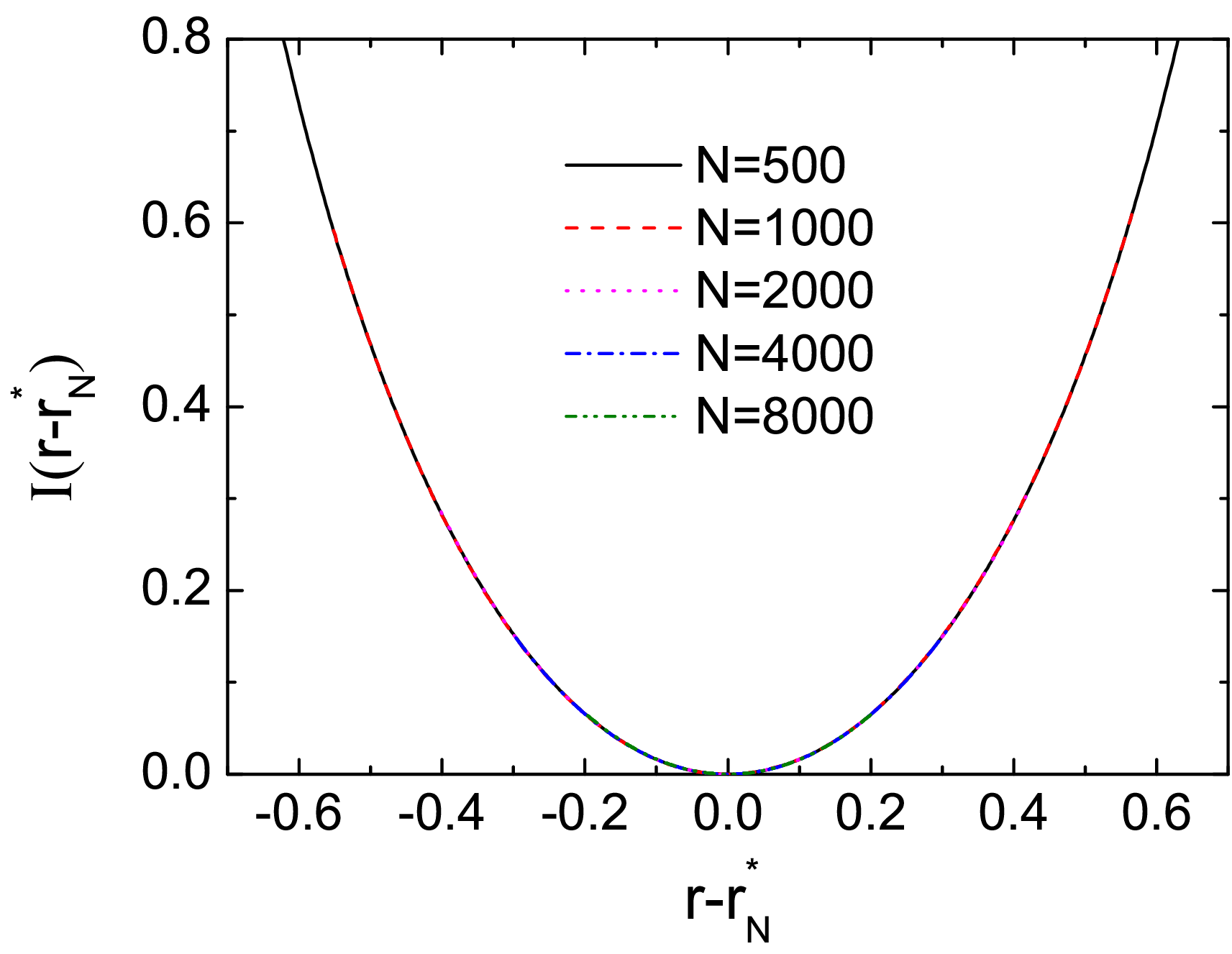}}
\caption{(color online). Large deviation rate function $I(r-r_N^*)$
on Poisson random graphs with average degree $\left\langle k
\right\rangle=6$. \label{fig3}}
\end{figure}

We first consider the Poisson random graphs whose degree
distribution follows $P\left( k \right) = {{{e^{ - \left\langle k
\right\rangle }}{{\left\langle k \right\rangle }^k}} \mathord{\left/
 {\vphantom {{{e^{ - \left\langle k \right\rangle }}{{\left\langle k \right\rangle }^k}} {k!}}} \right.
 \kern-\nulldelimiterspace} {k!}}$ with average degree ${\left\langle k \right\rangle
}=6$. In Fig. \ref{fig1}, we show the logarithm values of
$\rho_N(r)$ for several different $N$. Using the MHR method, the
probabilities as small as $e^{-200} \simeq 10^{-87}$ are easily
accessible. As $N$ increases, the width of the distribution of
$\rho_N(r)$ becomes narrower. The typical value $r_N^{*}$ of $r$,
i.e. the most probable value of $r$ corresponding to the maximum in
$\rho_N(r)$, is very close to zero. To investigate the size effect
of $\rho_N(r)$ in more detail, we have computed the expected value
$\left\langle r \right\rangle $ and the variance of $\sigma _r^2$ of
$r$ as a function of $N$. We find that $\left\langle r \right\rangle
$ is always negative for all the $N$'s and decays in magnitude with
$N$. As shown in Fig. \ref{fig2}(a), the minus $\left\langle r
\right\rangle $ can be well fitted linearly with $N$ in the log-log
plot, $ - \left\langle r \right\rangle \sim {N^{ - \nu }}$, with the
exponent $\nu=1.08$. In Fig. \ref{fig2}(b), we show that $\sigma
_r^2$ decreases with $N$ in a power-law way as well, $\sigma _r^2
\sim {N^{ - \xi }}$, with the exponent $\xi=0.99$ that is very close
to one. This implies that the fluctuation of $r$ on Poisson random
graphs is inversely proportional to the system size $N$, in
accordance with the central limit theorem.

Next, we want to check whether the $\rho_N(r)$ obeys a large
deviation principle. To \textcolor{blue}{that} end, we first make a shift $r_N^*$ in
$r$ such that the locations of the maximum in $\rho_N(r-r_N^*)$
coincide for all the $N$'s. We then scale the logarithm of $
\rho_N(r-r_N^*)$'s with $N^{-\xi}$ providing that the $\rho_N(r)$
obeys a Gaussian form around $r=r_N^*$. Thus, we suggest a form of
$\rho \left( r- r_N^* \right) \asymp {e^{ - {N^\xi }I\left( {r-
r_N^* } \right)}}$, where $I$ is the large deviation rate function
that is convex and possesses its unique minimum at $r= r_N^*$. \textcolor{blue}{Finally, we make a shift on $I$ so that $I_{\min}=0$ at $r= r_N^*$, which is often done because only $I_{\min}=0$ makes sense for $N \to \infty$.} This suggestion is verified in Fig. \ref{fig3}, in which one can see that
all the curves for each $N$ coincide not only near $r_N^*$, but also
far from $r_N^*$.

\section{Scale-free networks}

\begin{figure*}
\centerline{\includegraphics*[width=1.8\columnwidth]{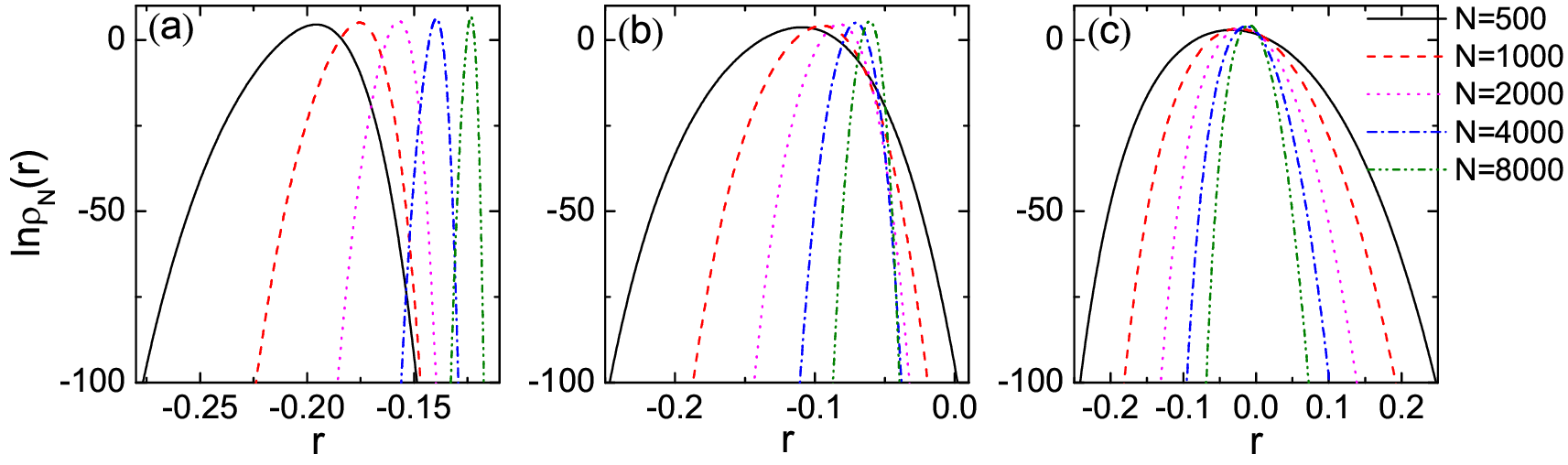}}
\caption{(color online). Logarithm of the probability distribution
$\rho_N(r)$ of the assortativity coefficient $r$ on scale-free
networks with different network size $N$ and a fixed minimal degree
$k_0=3$. The power-law exponent of degree distribution is
$\gamma=2.3$ (a), $2.5$ (b), and $3.0$ (c). \label{fig4}}
\end{figure*}

\begin{figure}
\centerline{\includegraphics*[width=1.0\columnwidth]{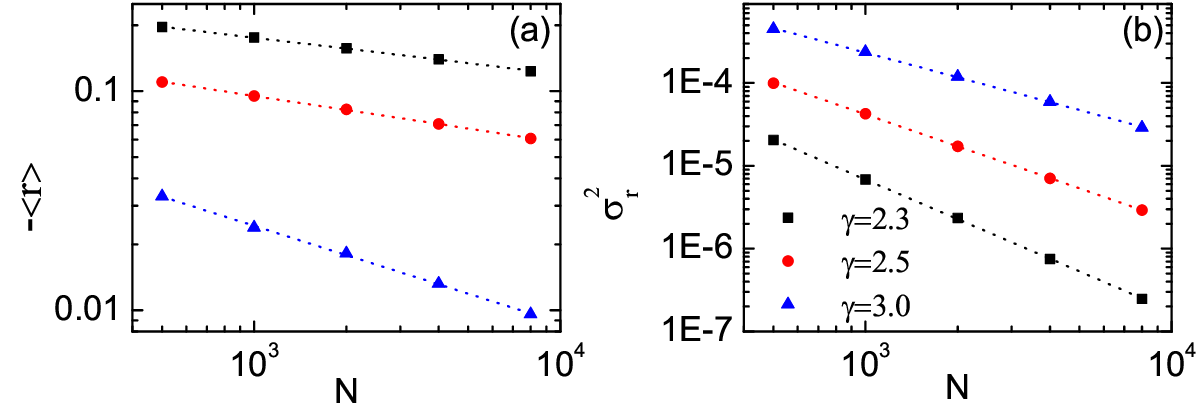}}
\caption{(color online). Log-log plot of the minus expected value
$-\left\langle r \right\rangle $ (a) and the variance $\sigma _r^2$
(b) of $r$ as a function of $N$ on scale-free networks with the
minimal degree $k_0=3$. The dotted lines show the linear fittings.
\label{fig5}}
\end{figure}

We now consider the case of scale-free networks whose degree
distribution follows a power-law function, \textcolor{blue}{$P(k) =(\gamma-1)k_0^{\gamma-1} k^{-\gamma}$, where $k_0$ is the
minimal degree, and $\gamma$ is degree distribution exponent. Here we focus on the range $2<\gamma<4$. The maximal degree
$k_{\max}$ is chosen by a natural cutoff, ${k_{\max }} = \min \left(
{{k_0}{N^{1/(\gamma  - 1)}},N - 1} \right)$ such that $\int_{k_0}^{\infty} P(k) \mathrm{d}k=1/N$.} In Fig. \ref{fig4}, we shows the logarithm of
$\rho_N(r)$ for three different $\gamma=2.3$ (a), 2.5 (b), 3.0 (c)
and for five different $N$'s. It can easily seen that for all cases
$\rho_N(r)$ are always unimodal. All the expected value of $r$ are
negative, $\left\langle r \right\rangle<0 $. This is especially
obvious for smaller $\gamma$. With the increment of $N$,
$\left\langle r \right\rangle $ moves to zero gradually. In Fig.
\ref{fig5}(a), we show that $\left\langle r \right\rangle $ can be
well fitted by the form of $ - \left\langle r \right\rangle \sim
{N^{ - \nu }}$. The exponent $\nu$ is dependent on $\gamma$, which
is $\nu=0.167$, 0.214, and 0.443 for $\gamma=2.3$, 2.5, and 3.0,
respectively. The fluctuations of $r$, $\sigma _r^2$, obey the
scaling law as well, $\sigma _r^2 \sim {N^{ - \xi }}$, as shown in
Fig. \ref{fig5}(b). The exponent $\xi$ decreases as $\gamma$
increases, which is $\xi=1.59$, 1.28, and 0.99 for $\gamma=2.3$,
2.5, and 3.0, respectively. That is to say, for highly heterogeneous
networks, they exhibit anomalously small fluctuations in $r$, since
$\xi>1$ implies that the fluctuations decay with $N$ faster than the
standard $1/N$ scaling.

\begin{figure*}
\centerline{\includegraphics*[width=1.8\columnwidth]{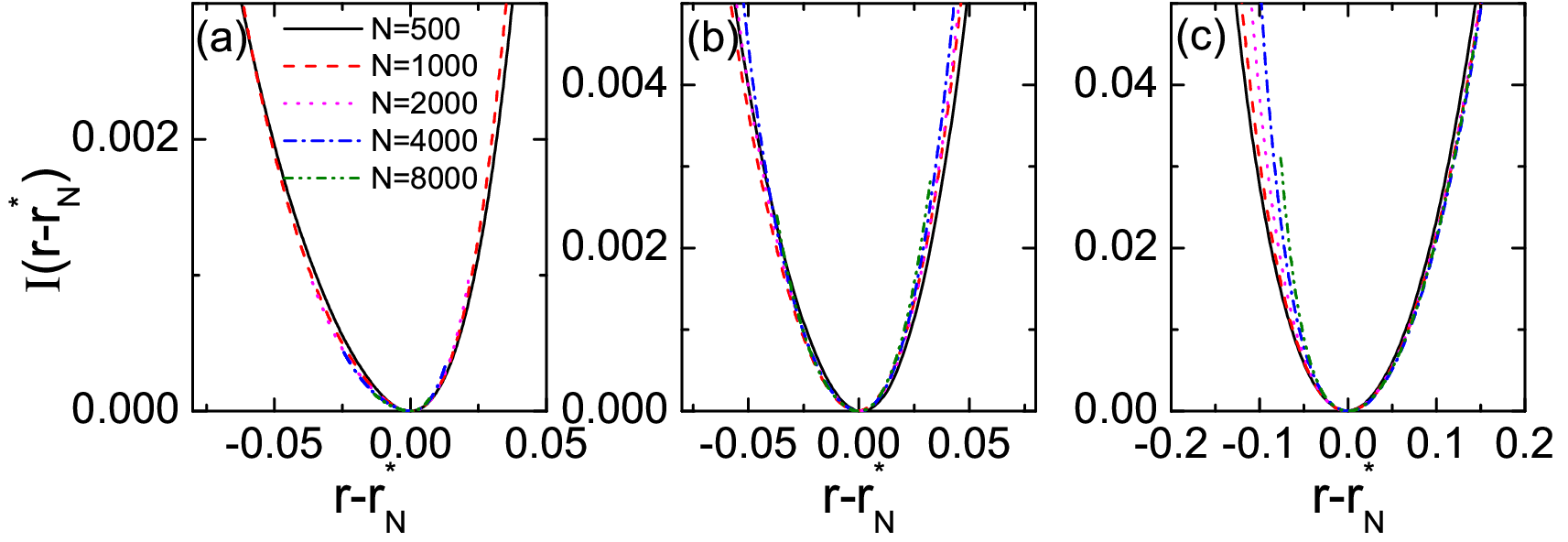}}
\caption{(color online). Large deviation rate function $I(r-r_N^*)$
on scale-free networks with the minimal degree $k_0=3$. The exponent
of degree distribution is $\gamma=2.3$ (a), $2.5$ (b), and $3.0$
(c). \label{fig6}}
\end{figure*}

In Fig. \ref{fig6}, we show the large deviation functions for
scale-free networks. As mentioned before, the large deviation
functions are obtained by $I \asymp - { {N^{-\xi}} } \ln \rho \left(
r- r_N^* \right)$. As expected, all the data coincide for different
$N$.

\section{Configuration network model with soft constraints}
\textcolor{blue}{Finally}, we shall compare the scaling behavior of the assortativity
coefficient $r$ between two different ensembles of configuration
model. The first one, as we studied before, is microcanonical, in
which degree sequence $\{k_1, \cdots, k_N\}$ are fixed. The second
one is canonical ensemble that is easier to handle mathematically,
and it is called the exponential random graph model in network
science
\cite{PhysRevE.70.066117,Garlaschelli_NJP2011,Garlaschelli_NatRevPhys2019}.
In the canonical ensemble, the hard constraints in microcanonical
ensemble are softened by enforcing only as expected values, i.e.
$\left\langle {{k_i}} \right\rangle  = {\bar k_i}$ for $i=1,\cdots,
N$. The canonical probability of a graph ${G}$ is written as
\cite{PhysRevE.70.066117,Garlaschelli_NJP2011,Garlaschelli_NatRevPhys2019,PhysRevE.87.062806,PhysRevE.93.062311,PhysRevE.104.014147}
\begin{eqnarray}
P\left( G \right) = \frac{1}{Z}{e^{ - H\left( G
\right)}},\label{eq9}
\end{eqnarray}
where $H$ is the graph Hamiltonian defined as
\begin{eqnarray}
H\left( G \right) = \sum\limits_i {{\theta _i}{k_i}\left( G \right)}
= \sum\limits_{i < j} {\left( {{\theta _i} + {\theta _j}} \right)}
{A_{ij}},\label{eq10}
\end{eqnarray}
and the normalizing quantity $Z$ is partition function that can be
calculated exactly,
\begin{eqnarray}
Z = \sum\limits_G {{e^{ - H\left( G \right)}} = } \prod\limits_{i <
j} {\left( {1 + {e^{ - {\theta _i} - {\theta _j}}}}
\right)}.\label{eq11}
\end{eqnarray}
Substituting Eq. (\ref{eq10}) and Eq. (\ref{eq11}) into Eq.
(\ref{eq9}), $P(G)$ can be written as the mass probability function
of a Bernoulli-distributed binary random variable $A_{ij}$
(adjacency maxtrix),
\begin{eqnarray}
P\left( G \right) = \prod\limits_{i < j} {p_{ij}^{{A_{ij}}}} {\left(
{1 - {p_{ij}}} \right)^{1 - {A_{ij}}}},\label{eq12}
\end{eqnarray}
with success probability
\begin{eqnarray}
{p_{ij}} = \frac{{{x_i}{x_j}}}{{1 + {x_i}{x_j}}},\label{eq13}
\end{eqnarray}
where $x_i=e^{-\theta_i}$ is called fugacity that can be obtained
numerically by solving constraint equations,
\begin{eqnarray}
\left\langle {{k_i}} \right\rangle  = \sum\limits_{j \ne i}
{{p_{ij}}}  = \sum\limits_{j \ne i} {\frac{{{x_i}{x_j}}}{{1 +
{x_i}{x_j}}}} .\label{eq14}
\end{eqnarray}

\begin{figure}
\centerline{\includegraphics*[width=1.0\columnwidth]{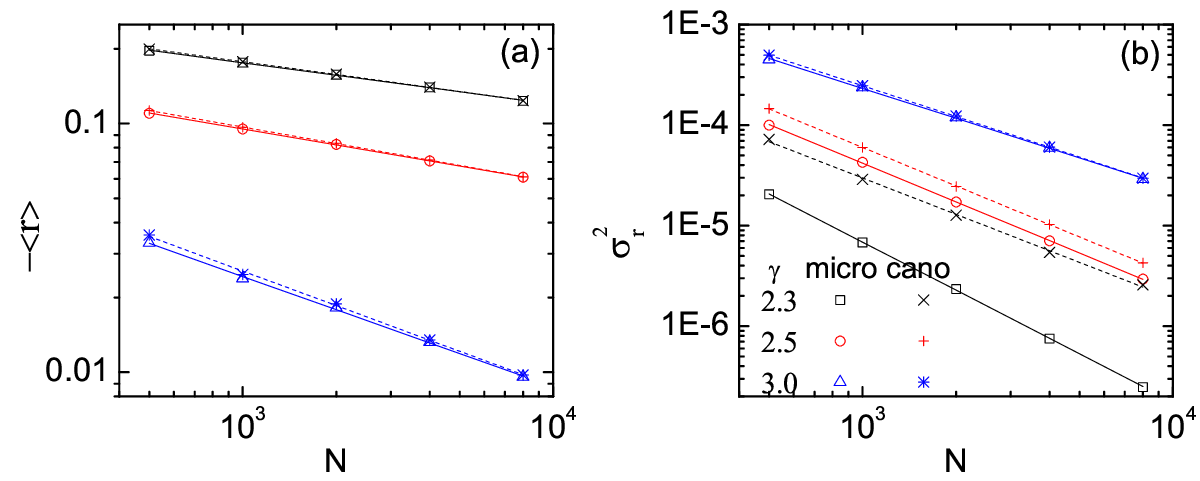}}
\caption{(color online). Comparison for the minus expected value
$-\left\langle r \right\rangle $ (a) and the variance $\sigma _r^2$
(b) of $r$ as a function of $N$ between the microcanonical ensemble and the
canonical ensemble. The solid lines and dotted lines show the
linear fittings for the microcanonical ensemble and canonical
ensemble, respectively. \label{fig7}}
\end{figure}

In Fig. \ref{fig7}, we compare the results of canonical scale-free
model with those of the microcanonical scale-free model for three
different values of $\gamma$. \textcolor{blue}{For each $\gamma$ and each $N$, we generate at
least 5000 realizations of canonical configuration networks
according to Eq. (\ref{eq13}) to obtain mean value and variance of $r$}, in which the expected values of node
degrees are the same as the degree sequence in microcanonical
configuration networks.  In canonical model, one can see that both
$-\left\langle r \right\rangle $ and $\sigma_r^2$ decay with
power-law as $N$ increases. On the one hand, the values of
$\left\langle r \right\rangle $ are almost independent of specific
ensemble and share the same scaling exponent $\nu$. On the other
hand, the values of $\sigma_r^2$ in canonical model are always
larger than those in microcanonical model. This is especially
obvious for smaller values of $\gamma$. The result is as expected
because in canonical model the degree of each node is fluctuating
from one network realization to another. For $\gamma=2.5$ and
$\gamma=3$, the scaling exponents $\xi$ are almost the same in the
two ensembles. However, for $\gamma=2.3$, $\xi \simeq 1.2$ in
canonical model is less than 1.59 in the microcanonical model.

\begin{figure}
\centerline{\includegraphics*[width=1.0\columnwidth]{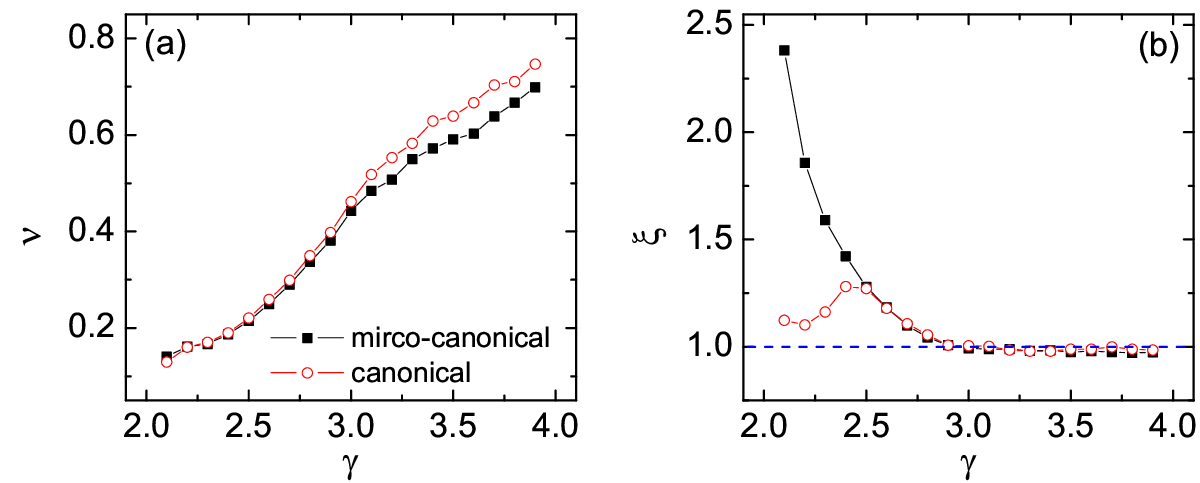}}
\caption{(color online). Scaling exponents $\nu$ (a) and $\xi$ (b)
as a function of degree distribution exponent $\gamma$ in scale-free
networks for the microcanonical ensemble and canonical
ensemble. The dashed line in (b) is only for eye-guide purpose.
\label{fig8}}
\end{figure}

In Fig. \ref{fig8}(a) and Fig. \ref{fig8}(b), we show the scaling
exponents $\nu$ and $\xi$ as a function of $\gamma$, respectively.
In the two ensembles $\nu$ increases monotonically as $\gamma$
increases. When $\gamma<3$, $\nu$ are almost the same, and when
$\gamma>3$, $\nu$ in canonical ensemble is slightly larger. However,
$\xi$ changes with $\gamma$ in two different trends. When
$\gamma>2.5$, $\xi$ in the two ensembles are almost the same, and
remains constant around one when $\gamma>3$. For $\gamma<2.5$, $\xi$
in microcanonical model are obviously larger than those in canonical
model. For example, for $\gamma=2.1$ we have $\xi=2.38$ in
microcanonical model and $\xi=1.22$ in canonical model. \textcolor{blue}{From Fig. \ref{fig8}(b), one can see that when $\gamma>3$ the scale-free networks start to share the same scaling exponents as the Poisson-distributed random graphs. Intuitively, it seems to be relevant to the divergence of the second moment of the degree distribution on scale-free networks with $\gamma>3$. It may be hopeful to establish this possible connection in the exponential random graph models as it is easier to handle mathematically in the canonical ensemble.} \textcolor{blue}{We have realized that in a recent paper \cite{PhysRevE.104.014147}, the authors used the two-star model \cite{PhysRevE.70.066146} to study degree correlations between the nearest and next nearest neighboring nodes. They analytically calculated the degree assortativities and showed that they are nonmonotonic functions of the model parameters, with a discontinuous behavior at a first-order transition. However, in the work the authors did not observe a broad degree distribution such as power law form that are properties of many empirical networks. Therefore, it is still a challenging problem. }

\section{Conclusions}
In summary, we have used the MHR method to obtain the probability
distribution $\rho_N$ of the assortativity coefficient $r$ on
configuration networks. This method enable us to obtain the
rare-probability tails of $\rho_N(r)$ within the allowable
computational time. We show that $\rho_N(r)$ satisfies a large
deviation principle after a shift $r_N^*$ in $r$, $\rho_N \left( r-
r_N^* \right) \asymp {e^{ - {N^\xi }I\left( {r- r_N^* } \right)}}$,
in which $I(r)$ is the large deviation rate function that is convex
and possesses its unique minimum at $r=r_N^*$. We find that $\xi=1$
in Poisson random graphs and scale-free networks with $\gamma>3$,
indicating a normal fluctuations scaling of $r$ with $N$ in such
networks, $\sigma_r^2 \propto 1/N$. Interestingly, $\xi>1$ for
$\gamma<3$, showing an anomalously fast decay in the fluctuation of
$r$ as $N$ increases. Such an anomalous phenomenon in time-consuming
observables have also been found in some other systems
\cite{PhysRevLett.113.078101,Majumdar_EPL_2016,Meerson_JSTAT_2017,PhysRevLett.121.060201,PhysRevLett.121.090602}.
Furthermore, we show that in the canonical ensemble $\xi$ is slightly greater than
one for $\gamma<2.5$ but is obviously less than that in the
microcanonical model. This suggests that the \textcolor{blue}{anomaly} in
fluctuations of $r$ is not very significant in the canonical ensemble.

\textcolor{blue}{In the future, it is worthy investigating the joint distribution of assortativity coefficient $r$ and other topological observables, such as the average shortest path length, the largest eigenvalue of adjacency matrix or the second smallest eigenvalue of the Laplacian matrix, using the MHR method. This will surely deepen the understanding of the role of degree assortativity on dynamical precesses on configuration networks \cite{PhysRevLett.89.108701,PhysRevE.66.047104,PhysRevLett.90.028701,Sinha2005,PhysRevE.75.046113,PhysicaD224.123.2006}.}

Recently, we have noticed that large deviation theory has been used
to uncover atypical structural and dynamical characteristics of
complex networks, such as a first-order percolation transition
subject to a rare initial damage
\cite{PhysRevE.97.022314,Bianconi_JSTAT_2019}, a first-order phase
transition in the condensation of node degrees
\cite{PhysRevE.100.012305}, localization transitions \cite{JPA49.184003.2016,PhysRevE.99.022137,Espigares2021} and optimal paths \cite{PhysRevE.103.022319} of dynamical
observables in random walk model
, and epidemic extinction
\cite{PhysRevLett.117.028302,PhysRevE.95.052317,PhysRevLett.123.068301}
and spin model \cite{PhysRevLett.123.068301,Chaos27.081102}. In the
future, we believe that large deviation theory and related
rare-event simulation methods may inspire more research works in
network science.


\begin{acknowledgments}
We acknowledge supports from the National Natural Science Foundation
of China (Grant Nos. 11875069, 11975025, 12011530158, 61973001) and the
Key Scientific Research Fund of Anhui Provincial Education
Department (Grant No. KJ2019A0781))
\end{acknowledgments}

%

\end{document}